\documentclass[preprint,prb,amsfonts,showpacs]{revtex4}
\usepackage[dvips]{graphicx}% Include figure files

\usepackage{bm}% bold math
\usepackage{color}
\begin{document}

\title{Thermoelectricity in Nanowires: A Generic Model} 
\author{Shadyar Farhangfar} 
\email{shadyar.farhangfar@gmail.com}
\affiliation{Institute of Applied Physics, University of Hamburg, Jungiusstrasse 11, D-20355 Hamburg, Germany} 
\date{\today}
\begin{abstract}
By employing a Boltzmann transport equation and using an energy and size dependent relaxation time ($\tau$) approximation (RTA), we evaluate self-consistently the thermoelectric figure-of-merit $ZT$ of a quantum wire with rectangular cross-section. The inferred  $ZT$  shows  abrupt enhancement in comparison to its counterparts in bulk systems. Still,  the estimated $ZT$  for the representative  Bi$_2$Te$_3$ nanowires and its dependence on wire parameters deviate considerably from those predicted by the existing RTA models with a constant $\tau$. In addition,  we address contribution of the higher energy subbands to the transport phenomena, the effect of chemical potential tuning on $ZT$, and correlation  of $ZT$ with quantum size effects (QSEs). The obtained results are of general validity for a wide class of systems and may prove useful in  the ongoing development of the modern thermoelectric applications. 
\end{abstract}
%%%%%%%%%%%%%%%%%%%%%%%  
%73.23.-b Electronic transport in mesoscopic systems  
%73.43.Qt Magnetoresistance 
%73.50.-h Electronic transport phenomena in thin films (for electronic transport in mesoscopic systems, see %%@
%73.23.-b; see %also 73.40.-c Electronic transport in interface structures; for electronic transport in %%@
%nanoscale%materials and structures, %see %73.63.-b) 
%73.50.Jt Galvanomagnetic and other magnetotransport effects (including thermomagnetic effects)  
%73.63.-b Electronic transport in nanoscale materials and structures (see also 73.23.-b Electronic transport in %%@
%mesoscopic systems) 
%73.63.Nm Quantum wires 
%73.50.Lw Thermoelectric effects 
%73.63.Nm Quantum wires  
%%%%%%%%%%%%%%%%%%%%%%%%%%%%%%%  

\maketitle

\section{Introduction}

Thermoelectricity deals with conversion of heat to electricity and vice versa. Last few years, mainly due to  the recent advances in material science and fabrication techniques on the one hand and a quest for alternative sources of energy generation on the other hand,  have witnessed a rejuvenation of interest in the  thermoelectric phenomena  \cite{disalvo99}-\cite{kim09}.  Performance of a thermoelectric device is defined through its dimensionless figure-of-merit $ZT\equiv{S^2}{\sigma}T/\kappa$. Here, $T$ is the absolute temperature, $S$ stands for the thermopower (the Seebeck coefficient), and $\sigma$ and $\kappa$ represent the  electrical and thermal conductivities, respectively. At first glance, a large $ZT$ might be achievable  by maximizing $\sigma$ and $S$, and by minimizing $\kappa$. The problem, however, lies in the fact that these quantities are interdependent and each of them is rather sensitive to the  material properties, to the  temperature, and to the size and geometry of the underlying  building blocks in a miniaturized thermoelectric module. Consequently, often one of the two not-completely-independent strategies is exploited: increasing of the power factor $P\equiv{S^2}{\sigma}$ which is largely an electronic property, or decreasing of $\kappa$ which consists of  electronic $\kappa_e$ and lattice $\kappa_{ph.}$ contributions by manipulation of the latter. A large $P$ can be achieved by tuning of the chemical potential level  through doping or external electric fields (gating) and by having control over the scattering mechanisms of the charge carriers in the device. Reduction in $\kappa_{ph.}$, instead, can be obtained by exploitation of material systems built up from  heavy elements and by  intentional enhancement of the phonon scattering events through introduction of lattice imperfections (impurity atoms, superlattice structures, \textit{etc.}) or by intensifying of the surface scattering of the heat carriers from the specimen boundaries in the systems with reduced dimensionality.

In what follows, assuming a constant $\kappa_{ph.}$, we focus on the charge carrier contributions to the thermoelectric properties of a one-dimensional (1D) wire with discrete energy levels. To demonstrate the applicability of the present approach, we evaluate $ZT$ in bismuth telluride nanowires and discuss the obtained results with an eye on the predictions of the earlier models. 
%%%%%%%%%%%%%%%%%%%%%%%%%%%%%%%%%%%%%%%%%%%%%%%%%%%%%%%   
\section{Theoretical model}

Current generic models (as compared to the \textit{ab initio} calculations) for evaluation of the thermoelectric figure-of-merit  are mostly based on the solution of Boltzmann transport equation (BTE) with various degrees of sophistications. An efficient approximation for  solving BTE can be attained by its linearization and through introduction of a relaxation time $\tau$, the time period within which the system gains its equilibrium after removal of the external stimulus \cite{ziman01}. Consequently,  the components of the transport tensors of the system  
$[{\cal L}^{(l)}]_{\alpha\delta}$ can be obtained through the relation  
%%%%%%%%%%%%%%%%%%%%%%%%%%%%   
\begin{eqnarray}
\label{basicLmn}
[{\cal L}^{(l)}]_{\alpha\delta}=\frac{\nu{q^{2-l}}}{2\pi}\int(-{{\partial}_{E} f_{0}})
{\tau({\bf k})}({E}({\bf k})-{\zeta
})^{l}{\bf v}_{\alpha}
{\bf v}_{\delta}{\rm d}{\bf k},
\end{eqnarray}
%%%%%%%%%%%%%%%%%%%%%%%%%%%%%%%%%%%%%%    
where $\sigma\equiv{{\cal L}^{(0)}}$ designates the electrical conductivity,  
$S\equiv\frac{1}{T}\frac{{\cal L}^{(1)}}{{{\cal L}^{(0)}}}$ is the Seebeck coefficient, and 
$\kappa_{e}\equiv\frac{1}{T}[{\cal L}^{(2)}-\frac{{{\cal L}^{(1)}}{\cal L'}^{(1)}}{{\cal L}^{(0)}}]$ is the electronic contribution to the thermal conductivity \cite{ziman01}. Above, $q$ is the electrical carrier charge, 
${E}({\bf k})$ is the energy-wavevector dispersion relation,  $\hbar{\bf v}({\bf k})\equiv{\nabla}_{\bf k}{E}({\bf k})$ defines the velocity operator ($\hbar$ is the reduced Planck constant), and $\nu$ is the valley degeneracy. The equilibrium Fermi-Dirac distribution function  is given by $f_{0}({E}){\equiv}[1+\exp\beta({E}-{\zeta})]^{-1}$, where $\beta\equiv1/k_{B}T$  ($k_B$ is the Boltzmann constant) and $\zeta$ denotes the chemical potential.

The precise form of the dispersion relation above can only be derived through detailed band structure calculations and  depends strongly on the material properties and on the boundary conditions in the system of interest. Similarly, the relaxation time for any specific process depends on the  density of energy states (DOS) and on the types of the scattering mechanisms taken into account. In most calculations, however, one expresses DOS  via the single-particle eigenenergies and assumes a power-law dependence of the relaxation time on energy, $\tau({E})\propto{E}^{\alpha}$. It was only under such conditions that A. F. Ioffe arrived at his compact expressions in terms of the  Fermi-Dirac integrals for the evaluation of the transport coefficients in Eq. (\ref{basicLmn}) above \cite{ioffe57,goldsmid01}.  Despite this, such basic assumptions have been widely overlooked in the latter evaluations of the $ZT$ values  in low-dimensional systems, leading ultimately to the estimations of unrealistically high, monotonously size-dependent,  $ZT$'s in atomically thin  nanowires \cite{hicks93I,hicks93II,sun99,handbook06}. It is worth reminding that in the limit of ultra narrow nanowires, the whole concept of a size-independent DOS and any argument subsequent to it will collapse. In such cases, one has to resort to the other proper techniques like, \textit{e.g.}, tight-binding models and check  for the validity of the obtained results self-consistently. Furthermore, as shown  elsewhere \cite{mahan96}, even minor variations in the form of DOS [and subsequent changes in  $\tau({E})$] will affect $ZT$ values dramatically. It is also elucidating to remind that, based on a purely mathematical analysis of the functional in Eq. (\ref{basicLmn}),  Mahan and Sofo estimate in \cite{mahan96}  a universal (that is, independent of size, temperature, and specific material properties) finite upper limit for $ZT$ in semiconductor thermoelectric materials.   Based on these facts and to circumvent the above-mentioned crucial shortcomings in the existing theoretical models  for   low-dimensional thermoelectric systems,  starting with the basic principles, below we obtain  the transport coefficients and $ZT$ values specifically for a 1D  nanowire with quantized energy levels. The key issue in such a treatment is the derivation of a proper, \textit{i.e.} size and energy dependent, expression for the relaxation time.

Let us assume that $N$ scatterers each with a scattering strength $V_0$ 
are randomly distributed at positions ${\bf R}_j$ along the wire, 
$V({\bf r})={\sum}_{j=1}^{N}{V_0}\delta\left({\bf r}-{\bf R}_j\right)$,   and make use of the Fermi's golden rule 
${\tau}_{i\rightarrow{f}}^{-1}=({2\pi}/{\hbar}){\big|}
\langle {i\mid V({\bf r})\mid f}\rangle{\big|}^{2}$ to derive the corresponding scattering rates. By averaging over  the configuration of the scattering centers \cite{farhangfar06}, we arrive at        
%%%%%%%%%%%%%%%%%%%%%%%%%%%%%%%%%%%%%%%%    
\begin{eqnarray}
\label{tau}
{{\tau}_{mn\rightarrow m'n'}^{-1}}=\frac{\pi}{2}\frac{\varrho}{\hbar}{}
{V_0^2}{\Lambda}_{m'n'}^{mn}g(E)\cdot\Omega\big|_{{E}={E}(k'm'n')},  
\end{eqnarray}
%%%%%%%%%%%%%%%%%%%%%%%%%%%%%%%%%%%%%%%%%%% 
where  ${\Lambda}_{m'n'}^{mn}\equiv(2+{\delta}_{mm'})(2+{\delta}_{nn'})$ and $\varrho\equiv{N}/{\Omega}$. Here, $\Omega{\equiv}wtL$ is the volume of the wire;   $w$ and $t$ are the lateral dimensions and $L$ stands for the wire length.  The energy density of states per unit volume is given by  
%%%%%%%%%%%%%%%%%%%%%%%%%%%%%%%%%%%%%%%%%%%%%%%%%%%%%%%%   
\begin{eqnarray}%\nonumber
\label{dos}
g({E})=\frac{s}{2\pi \hbar}\frac{1}{wt}\sqrt{\frac{m_{z}}{2}}
\sum_{i,j} \frac{\Theta\left({E}-{E}_{ij}\right)}{\sqrt{{E}-{E}_{ij}}}.  
\end{eqnarray}
%%%%%%%%%%%%%%%%%%%%%%%%%%%%%%%%%%%%%%%%%%%%%%%%%%%%%%%%%%%%  
Here,  $m_z$ is the component of the charge carrier mass along the wire axis, $s$ is the spin degeneracy, and $\Theta$ designates the Heaviside step function.  Above, use has been made of the fact that the single-particle wave functions and the corresponding eigenenergies of a  carrier confined in lateral dimensions $x$ and $y$ and traveling freely along the $z$-axis are given by ${\Psi}_{ij}(k)=({2}/{\sqrt{\Omega}})\sin({i\pi  x}/{w})\sin({j\pi  y}/{t})\exp{\left({\bf i} kz\right)}$ and $E {\equiv} {E_{ij}+{({\hbar^2}{k^2}/2{m_z})}}$, respectively; $E_{ij}\equiv({\hbar^2}{\pi^2}/{2})({{i^2}/{w^2}{m_x}}+{{j^2}/{t^2}{m_y}})$ are the subband energies, $i,j\in\mathbb{N}$, and $k$ is the wavenumber.

In Eq. (\ref{tau}) above, as  in most cases the density  and strength  of scatterers are unknown and may vary from one individual wire to other, one can make use of the definition of carrier mobility $\mu{\equiv}q\left<\tau\right>/{m^\ast}$ and  account for the product ${\varrho}V_0^2$ through \cite{rhov2}     
%%%%%%%%%%%%%%%%%%%%%%%%%%%%%%%%  
\begin{eqnarray}%\nonumber
\label{rhoV0}
\frac{1}{\varrho{V_0^2}}=\frac{s\nu}{4}\frac{\Omega}{wt}\frac{{\mu}m^{\ast}}{q{\hbar^2}}
\sqrt{\frac{m_z}{2k_{B}T}}
\sum_{m,n}
\frac{1}{\left<{\tau}_{mn}\right>}, 
\end{eqnarray}
%%%%%%%%%%%%%%%%%%%%%%%%%%%%%%%%%%%%%%%%   
where  the lifetime of the state $(m,n)$ is defined through the relation      ${\tau_{mn}^{-1}}\equiv{\sum_{m'n'}}{\tau_{{mn}\rightarrow m'n'}^{-1}}$ and its expectation value as    $\left<{\tau}_{mn}\right>\equiv
{{\int}{{{\tau}_{mn}({\varepsilon}){\varepsilon}g({\varepsilon}){\partial_{\varepsilon} f}{\rm d}{\varepsilon}}}} / 
{{\int}{\varepsilon}g({\varepsilon}){\partial_{\varepsilon} f}{\rm d}{\varepsilon}}$.  Here,   $\varepsilon\equiv\beta{E}$ is the reduced energy and $m^\ast$ is the effective mass (for a 1D wire aligned along the $z$-axis, ${m^\ast}={m_z}$). Now, substituting for the corresponding expressions and making a coordinate transformation  where $\varepsilon$ and the reduced chemical potential  ${\zeta^\ast}\equiv\beta{\zeta}$  (for electrons) are measured  from the bottom of the conduction band  \cite{ziman01}, the transport coefficients in Eq. (\ref{basicLmn}) can be expressed as        %%%%%%%%%%%%%%%%%%%%%%%%%%%%%%%%%%%%%%%%%%%%        
\begin{eqnarray}%\nonumber
\label{reducedLmn}
[{\cal L}^{(l)}_{ij}]_{z}&=&
{\frac{s\nu}{wt}\frac{4\pi\mu}{{\beta^{l+0.5}}}\frac{m^\ast}{m_z}\sqrt{\frac{m_z}{2}}\frac{{q^{(1-l)}}}{\hbar}}\sum_{m,n}\frac{1}{\left<\tau_{mn}\right>}{\cal{F}}^{l}_{ij}, 
\end{eqnarray}
%%%%%%%%%%%%%%%%%%%%%%%%%%%%%%%%%%%%%%%%%%%%%% 
where ${\cal{F}}^{l}_{ij}\equiv{\cal{F}}^{l}_{ij}({\zeta^\ast})$ and 
%%%%%%%%%%%%%%%%%%%%%%%%%%%%%%%%%%%   
\begin{eqnarray}%\nonumber
\label{F}
{\cal F}^{l}_{ij}\equiv{\int_0^\infty}\text{sech}^{2}\left[\frac{{\varepsilon}+{\zeta^\ast}}{2}\right]
\tau_{ij}(\varepsilon)({\varepsilon}+{\zeta^\ast})^{l}({{\varepsilon}-{\varepsilon}_{ij}})^{1/2}\rm{d}{\varepsilon}.
\end{eqnarray}
%%%%%%%%%%%%%%%%%%%%%%%%%%%%%%%%%%%%%%%%       
Total contribution from all the subbands will now be ${\cal L}^{(l)}={\sum}_{i,j}{\cal L}^{(l)}_{ij}$. (The contribution of the holes can be accounted for analogously.)

This equation has to be compared to the corresponding ones based on a size-independent power-low  RTA approach introduced  originally in \cite{ioffe57} and exploited later by Dresselhaus and coworkers in their pivotal \cite{hicks93II} and subsequent studies \cite{sun99,handbook06} on low-dimensional thermoelectric systems.

\section{Results and their discussion}

Next, we evaluate $ZT$ for a bismuth telluride nanowire at $T=70$ K and at $300$ K. To make the comparison between the predictions of the present model and those of the previous studies \cite{hicks93II,sun99,handbook06} more transparent, we use the same material parameters, $m_x=0.32$, $m_y=0.08$, and $m^{\ast}=m_z=0.02$ for the electronic effective masses (all in units of the free electron mass), and take the same value for the mobility of electrons, $\mu_{z}=1200$ cm{$^2$}V{$^{-1}$}s$^{-1}$. The lattice thermal conductivity of bulk Bi$_2$Te$_3$ is 
$\kappa_{ph.}^{\rm bulk}\approx1.5$ W/Km.   The phonon confinement effects become considerable  only if the lateral dimensions of the wire, $w$ and $t$, are comparable in size to the phonon mean free path $\lambda_{ph.}$. As for bismuth telluride nanowires $\lambda_{ph.}\sim 1$ nm \cite{hicks93II}, for  wires with $w,t\gtrsim\lambda_{ph.}$, one can safely assume  $\kappa_{ph.}\approx\kappa_{ph.}^{\rm bulk}$     \cite{kappa}. 
%%%%%%%%%%%%%%%%%%%%%%%%%%%%%%%%%%%%%%%%%%%%%%%%%%%%%%   
%%%%%%%%%%%%%%%%%%%%%%%%%%%%%%%%%%%%%%%%%%%%%%%%%%%%%%%%%%%%%%%%%           
\begin{figure}\label{Fig.1}
\begin{center}
\includegraphics[width=120mm] {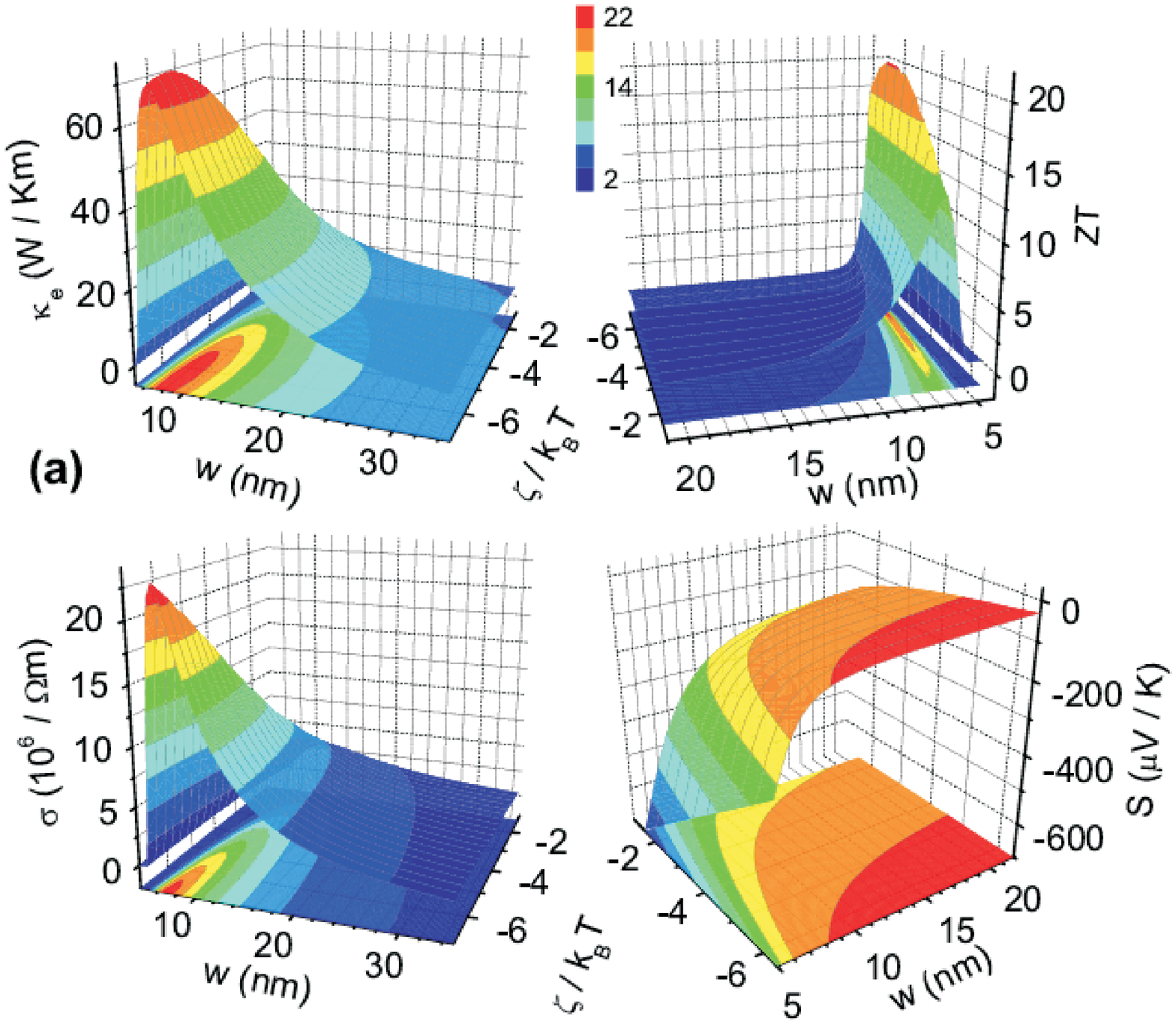} 
\includegraphics[width=120mm] {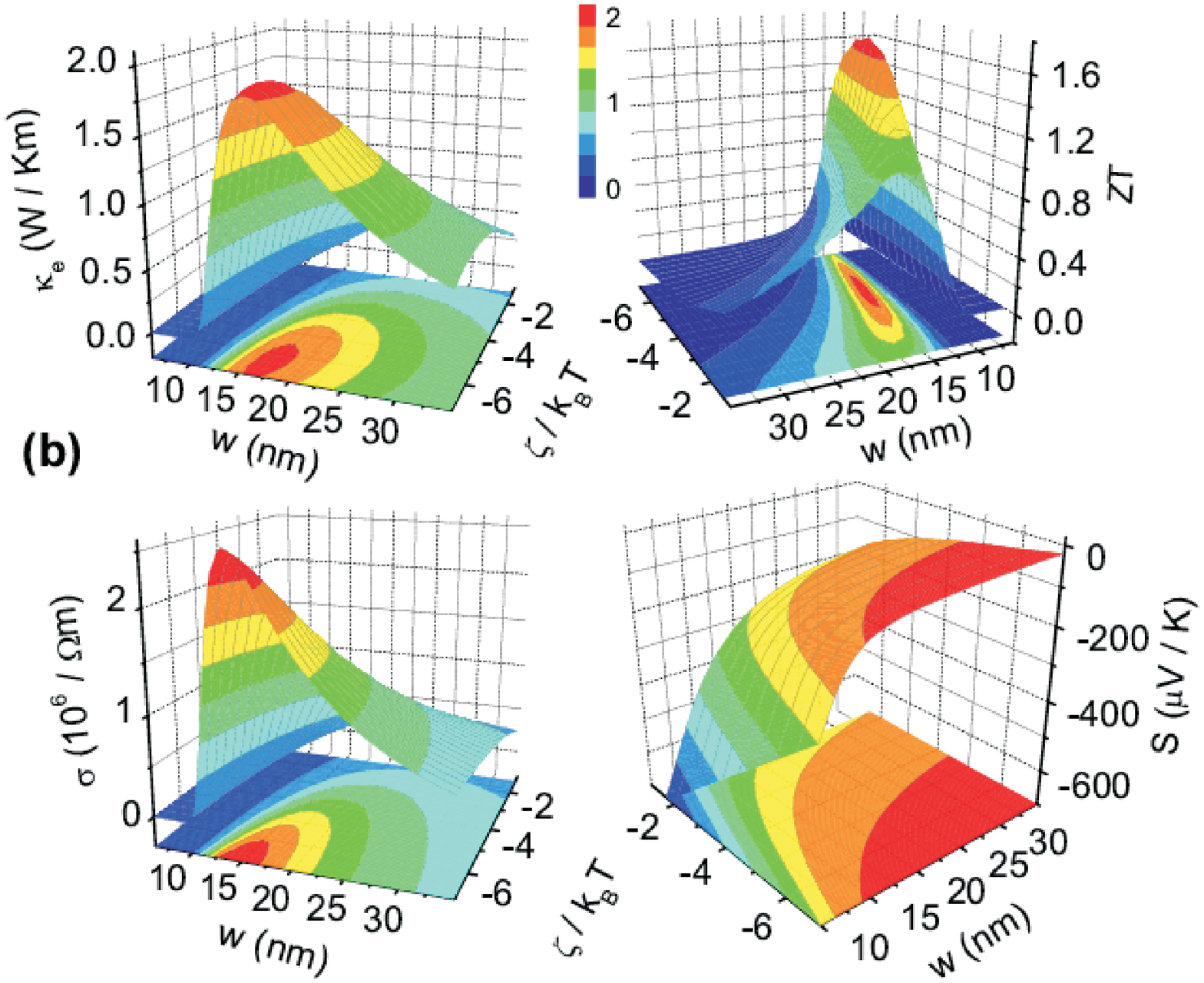}
\begin{caption}
{\textbf{(a)} Dependence of the transport properties and the thermoelectric figure-of-merit $ZT$ (only the electronic contribution) on wire width $w$ and on the chemical potential $\zeta$ (measured in units of thermal energy $k_{B}T$) for a square  Bi$_2$Te$_3$ nanowire at $T=300$ K. The valley degeneracy is $\nu=6$. Only the first subband  contribution is taken into account. \textbf{(b)} $T=70$ K. Notice the many fold decreases of the peak values and their shift  toward larger wire thicknesses.}
\end{caption}
\end{center}
\end{figure} 
%%%%%%%%%%%%%%%%%%%%%%%%%%%%%%%%%%%%%%%%%%%%%%%%%%%%%%%%%%%%%%%%%  
%%%%%%%%%%%%%%%%%%%%%%%%%%%%%%%%%%%%%%%%%%%%%%%%%%%%%%%%%%%%%%%%%              
\begin{figure}\label{Fig.2}
\begin{center} 
\includegraphics[width=120mm] {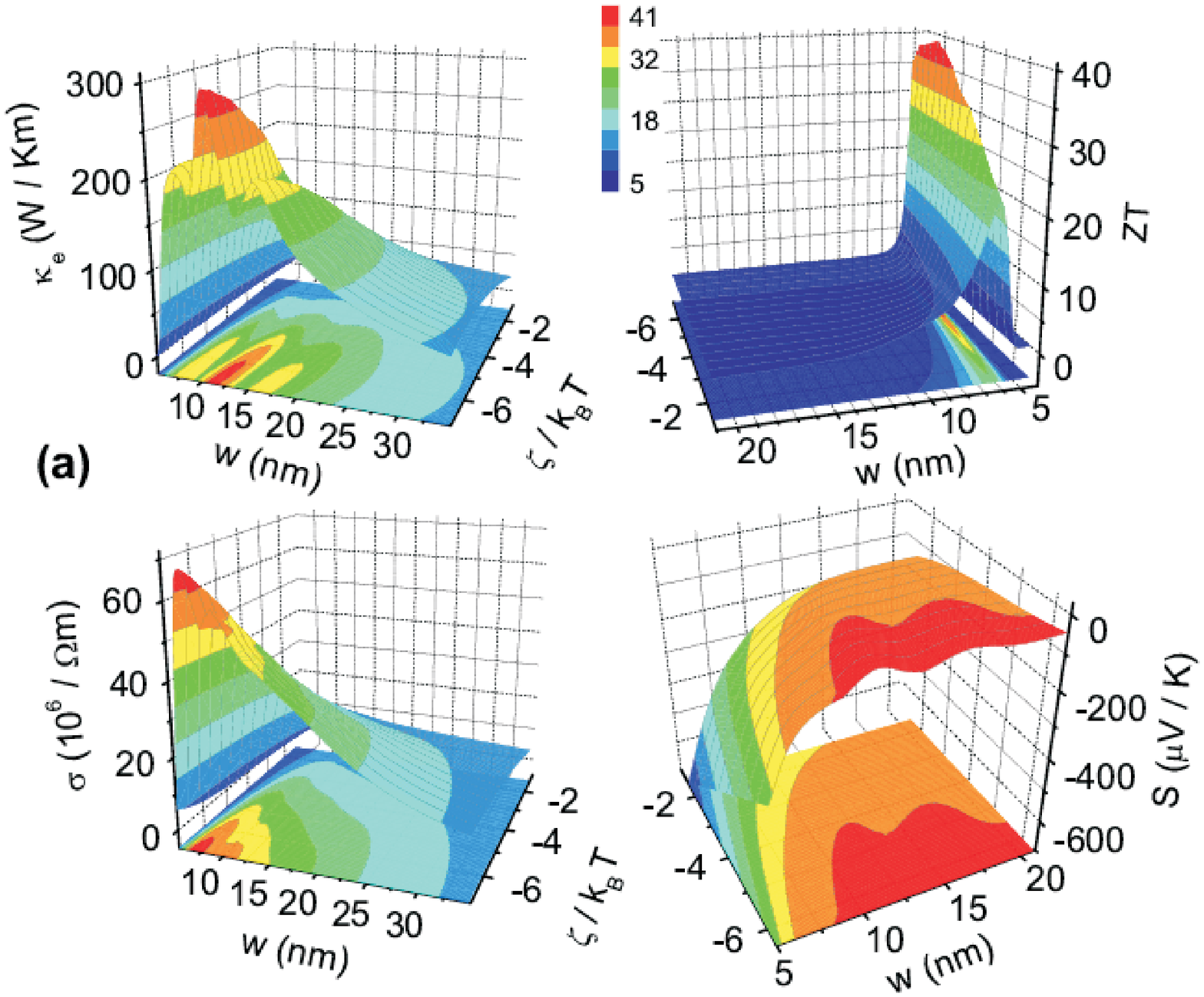}
\includegraphics[width=120mm] {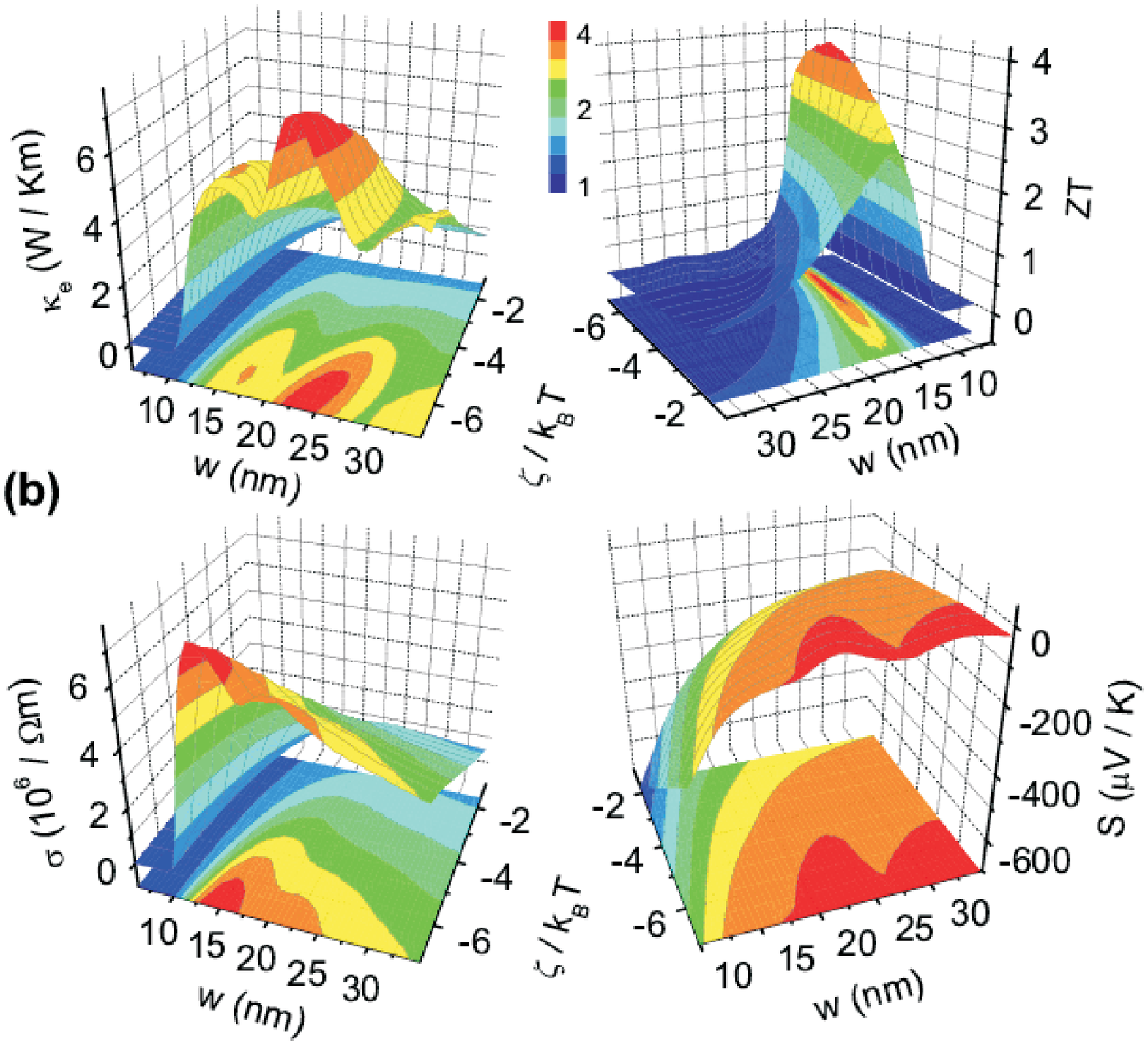}
\begin{caption}
{\textbf{(a)} Dependence of the  transport coefficients  on the wire size and on the chemical potential at $T=300$ K. Same parameters as in Fig. 1, except that here $(m,n)=(1,3)$. The clearly visible bumps in $\sigma$ and $\kappa_e$ are manifestations of the quantum size effects. \textbf{(b)} $T=70$ K.}
\end{caption}
\end{center}
\end{figure} 
%%%%%%%%%%%%%%%%%%%%%%%%%%%%%%%%%%%%%%%%%%%%%%%%%%%%%%%%%%%%%%%%%   
%%%%%%%%%%%%%%%%%%%%%%%%%%%%%%%%%%%%%%%%%%%%%%%%%%%%%%%%%%%%%%%%%%%                                           
Figure 1  shows dependence of the electronic contribution to the figure-of-merit of a  wire with square  cross-section on its size $w$ and on $\zeta^\ast$. Here, in contrast to the  constant-$\tau$ model of Hicks and Dresselhaus (HD) (see, the Appendix), dependence of $ZT$ on wire size is non-monotonous and after reaching its optimum at a certain width and chemical potential,  falls starkly to zero. This is a direct consequence of disappearance of electronic density of states for ultra thin wires. The most distinct differences between the two models are as follow: the peak values of the $\sigma$, $\kappa_e$, and $ZT$ in the present model show a shift toward larger wire thicknesses, $\sim 10$ nm, as compared to the plain  ${w^{-2}}$-dependence of the corresponding quantities predicted by the HD model; the thermopower $S$ given here is temperature dependent and reaches the overall high values  of the HD model only in very narrow wires; in contrast to the HD model, here the position of optimum $ZT$ in ${\zeta}w$-plane has a visible  dependence on temperature and, as temperature falls down, shifts  toward higher thicknesses; the asymptotic values of the transport coefficients for a  model square wire with $w=50$ nm given by the HD model are unrealistically smaller than the those obtained here (Table 1; see also the Table in Appendix).

To illustrate the effect of subband contributions and that of QSEs, the calculations were repeated for the first three subbands, up to $(i,j)=(1,3)$ (see, Fig. 2). Table 1 summarizes the results. Here, especially at $T=70$ K,  one can distinguish two clear maxima for $\sigma$, $\kappa_{e}$, and $S$ which arise from QSEs \cite{farhangfar06,farhangfar07} and which are smeared out in $ZT$. This observation is rather general for other subbands too and is a consequence of the fact that  $\sigma$ and  $\kappa_e$ have similar functional dependencies on the wire size and the QSEs distinguishable independently in each of them (and to a lesser extent in $S$) are compensated in $ZT$ by one another. Another interesting observation  is the abrupt enhancement of  $\kappa_{e}$ at thinner wire diameters and in the region where $\zeta^\ast$ approaches its optimum. Here, noticeably,  $\kappa_{e}\gg{\kappa}_{ph.}$. Alike, the HD model predicts an even stronger enhancement of $\kappa_e$ in  thinner wires (Appendix). Dependence of the thermal conductivity on dopant concentration (cf. tuning of $\zeta$), leading ultimately to the takeover of $\kappa_{ph.}$ by the electronic contribution $\kappa_e$, was observed in experiments with bismuth-antimony alloys some years ago \cite{sharp01}.  
%%%%%%%%%%%%%%%%%%%%%%%%%%%%%%%%%%%%%%%%%%%%%%%%%%%%%%%%%%%%%%%
\begin{center}
\begin{table}[h]
\begin{tabular}{ccccccc}
%% after \\: \hline or \cline{col1-col2} \cline{col3-col4} ...
$T ({\rm K})$ & $(i,j)$  & $\sigma(\rm{{10^6}/{\Omega}{m}})$ & $\kappa_{e}({\rm{W}/{Km}})$ 
& $S(\rm{{\mu}V/K})$  & $ZT({\nu=6})$ \\
%%(K) & (nm)& (GPa(nm)$^4$)& (GPa) & (nm) & \\
\hline\hline
  70 & HD  & $6.5\times10^{-5}$ & $4.8\times10^{-6}$ & -401.3  & $4.9\times10^{-4}$\\
  300 & HD  & $1.3\times10^{-4}$ & $4.3\times10^{-4}$ & -401.3  & $4.3\times10^{-3}$\\
\hline    
  70 & (1,1)  & 0.25 & 0.31& -92.9  & 0.08\\
  300 & (1,1)  & 0.53 & 2.96 & -85.6  & 0.27\\    
  70 & (1,2)  & 0.86 & 1.09 & -103.9 & 0.25\\
  300 & (1,2)  & 2.09 & 11.5 & -88.5  & 0.38\\  
  70 & (1,3)  & 1.45 & 2.10 & -105.4  & 0.31\\
  300 & (1,3)  & 4.55 & 24.8 & -92.3  & 0.44\\
\hline\\
\end{tabular}
\caption{Values  of the transport coefficients for a square $50$-nm-thick Bi$_2$Te$_3$ wire (see Text) with valley degeneracy $\nu=6$ at a fixed reduced chemical potential ${\zeta^\ast}=-3$. The indices $(i,j)$ correspond to the subband numbers up to which the electronic (vs. holes) contributions are taken into account ($\kappa_{ph.}=1.5$ W/Km). HD stands for the one-band  constant-$\tau$ model employed in previous studies  \cite{bejenari08,hicks93II,sun99,handbook06}. For more details, see the Table in Appendix.} 
\end{table}
\end{center}
%%%%%%%%%%%%%%%%%%%%%%%%%%%%%%%%%%%%%%%%%%%%%%%%%%%%%%%%%%%%
\section{Summary}

The approach described here, which is based on  the formulation of the scattering rates of the charge carriers in terms of an energy and size dependent expression for the density of states, can be readily extended to address transport and thermoelectric properties of a wide range of quantum systems subject to more sophisticated forms of the relevant DOS's. The model also holds potential to account for the non-diagonal contributions of the transport tensor elements to the kinetic properties. It can equally be  exploited in the metallic or semiconducting  regimes and it considers  inherently  the important issue of quantum size effects in low-dimensional structures.

This work is dedicated to the memory of Amirkhan Qezelli, the uncle, the childhood friend.

\section{Appendix} 

The complementary Table 2 outlines contributions of the higher energy subbands to the transport properties and to $ZT$ in a $50$-nm-thick Bi$_2$Te$_3$ nanowire with effective mass components ${m_x}=0.32$, ${m_y}=0.08$, and with ${m_z}=0.02$ along the wire axis (all in units of the free electron mass). We also summarize the main results of the constant-$\tau$ model and demonstrate its predictions in the appended Figure 3. 
%%%%%%%%%%%%%%%%%%%%%%%%%%%%%%%%%%%%%%%%%%%%%%%%%%%%%%%%%%%%%%%
\begin{center}
\begin{table}[h]
\label{Appendix Table}
\begin{tabular}{ccccccc}
$T ({\rm K})$ & $(m,n)$  &  $\sigma(\rm{{10^6}/{\Omega}{m}})$ & $\kappa_{e}({\rm{W}/Km})$ 
& $S(\rm{{\mu}V/K})$  & $ZT({\nu=1})$ & $ZT({\nu=6})$\\
%(K) & (nm)& (GPa(nm)$^4$)& (GPa) & (nm) & \\
\hline\hline
  70 & HD  & $1.1\times10^{-5}$ & $8.1\times10^{-6}$ & -401.3 & $8.1\times10^{-5}$ & $4.9\times10^{-4}$\\
  300 & HD  & $2.2\times10^{-5}$ & $7.2\times10^{-5}$ & -401.3 & $7.2\times10^{-4}$ & $4.3\times10^{-3}$\\
\hline 
  70 & (1,1)  & 0.041 & 0.052 & -92.9  & 0.016 & 0.082\\
  300 & (1,1)  & 0.089 & 0.494 & -86.0  & 0.099 & 0.265\\
  
  70 & (1,2)  & 0.144 & 0.182 & -103.9 & 0.065 & 0.252\\
  300 & (1,2)  & 0.348 & 1.92 & -88.5  & 0.240 & 0.377\\ 
  
  70 & (1,3)  & 0.242 & 0.350 & -105.5  & 0.102 & 0.314\\
  300 & (1,3)  & 0.758 & 4.13 & -92.3  & 0.344 & 0.442\\
  
  70 & (2,1)  & 0.160 & 0.206 & -95.9  & 0.061 & 0.228\\
  300 & (2,1)  & 0.354 & 1.96 & -86.6  & 0.230 & 0.360\\ 
  
  70 & (2,2)  & 0.552 & 0.691 & -107.8  & 0.205 & 0.477\\
  300 & (2,2)  & 1.39 & 7.62 & -89.1  & 0.362 & 0.420\\ 
   
  70 & (2,3)  & 0.908 & 1.31 & -109.7  & 0.273 & 0.492\\
  300 & (2,3)  & 3.02 & 16.41 & -93.0  & 0.438 & 0.471\\  
  
  70 & (3,1)  & 0.345 & 0.428 & -100.4  & 0.126 & 0.359\\
  300 & (3,1)  & 0.791 & 4.37 & -87.5 & 0.309 & 0.393\\ 
  
  70 & (3,2)  & 1.17 & 1.46 & -112.9  & 0.353 & 0.611\\
  300 & (3,2)  & 3.10 & 16.97 & -90.0  & 0.408 & 0.438\\
  
  70 & (3,3)  & 1.86 & 2.67 & -114.2  & 0.406 & 0.578\\
  300 & (3,3)  & 6.74 & 36.48 & -94.1  & 0.471 & 0.487\\
  \hline\\
\end{tabular}
\caption{Electrical conductivity $\sigma$, electronic thermal conductivity $\kappa_e$, the Seebeck coefficient $S$, and the  thermoelectric figure-of-merit $ZT$  for a quadratic Bi$_2$Te$_3$ nanowire (aligned along the $z$-axis) with $w=50$ nm   at a fixed chemical potential ${\zeta^\ast}=-3$ (measured from the bottom of the conduction band). The indices $(m,n)$ correspond to the subband numbers up to which the electronic (vs. holes) contributions are taken into account ($\kappa_{ph.}=1.5$ W/Km). HD represents the one-band  constant-$\tau$ model \cite{hicks93II,sun99,handbook06}. Note that with valley degeneracy $\nu=6$, $\sigma$ and $\kappa_e$ become simply six times larger while $S$ remains intact. Dependence of $ZT$ on $\nu$ is of the form  $ZT\propto{({\kappa_e}+{\kappa_{ph.}/\nu})^{-1}}$ and its values for $\nu=1$ and $\nu=6$ are given explicitly.} 
\end{table}
\end{center}
%%%%%%%%%%%%%%%%%%%%%%%%%%%%%%%%%%%%%%%%%%%%%%%%%%%%%%%%%%%%      
\begin{center}
\textbf{Constant relaxation-time model}
\end{center}
%\vspace{5mm}

Derivation of the transport properties in bulk systems, based on a constant relaxation-time approximation,  was originally given in \cite{ioffe57} and later discussed in detail by Nolas  \textit{et al.}  in   \cite{goldsmid01}. Later, extension of these results to the case of one-dimensional nanowires was obtained by Dresselhaus and coworkers \cite{hicks93II,sun99}. A more recent account is given in \cite{handbook06}. Below, we summarize the results.

The electronic (vs. holes) contribution to the transport coefficients $\sigma\equiv{{\cal L}^{(0)}}$ (electrical conductivity),   
$S\equiv-({1}/{qT}){{\cal L}^{(1)}}/{{\cal L}^{(0)}}$ (the Seebeck coefficient), and 
$\kappa_{e}\equiv({1}/{{q^2}T})[{\cal L}^{(2)}-{({{\cal L}^{(1)}})^2}/{{\cal L}^{(0)}}]$ (the electronic  thermal conductivity) are given by the following expressions. Here, $q$ is the magnitude of the elementary charge.  
%%%%%%%%%%%%%%%%%%%%%%%%%%%%%%%%%%%%%%%%%%%%        
\begin{eqnarray}
\label{HDcoefficients}
\nonumber
{\cal L}^{(0)}&=&\frac{1}{2}{D_e}F_{-1/2}    \\\nonumber
{\cal L}^{(1)}&=&\left({k_B}T\right){D_e}\left(\frac{3}{2}F_{1/2}-\frac{1}{2}{\zeta^\ast}F_{-1/2}\right)\\\nonumber 
{\cal L}^{(2)}&=&{({k_B}T)^2}{D_e}
\left(\frac{5}{2}F_{3/2}-3{\zeta^\ast}F_{1/2}+\frac{1}{2}{{\zeta^\ast}^2}F_{-1/2}\right),  
\end{eqnarray}
%%%%%%%%%%%%%%%%%%%%%%%%%%%%%%%%%%%%%%%%%%%%%% 
where 
%%%%%%%%%%%%%%%%%%%%%% 
\begin{equation}
\nonumber 
{D_e}\equiv{\nu}\frac{2q}{\pi{w^2}}{\left( \frac{2{k_B}T}{\hbar^2}\right)^{1/2}}{\sqrt{m^\ast}}\mu_{e}
\end{equation}    
%%%%%%%%%%%%%%%%%%%%%%%%%%%% 
and 
%%%%%%%%%%%%%%%%%%%%%%%%%%%%%   
\begin{equation}
\nonumber 
F_{i}{\equiv}F_{i}(\zeta^\ast)\equiv{\int_{0}^{\infty}}\frac{{\varepsilon^i}{\rm d}\varepsilon}{{\rm e}^{(\varepsilon-\zeta^\ast)}+1}.
\end{equation}
%%%%%%%%%%%%%%%%%%%%%%%%% 
%%%%%%%%%%%%%%%%%%%%%%%%%%%%%%%%%%%%%%%%%%%%%%%%%%%%%%%%%%%%%%%%%           
\begin{figure}
\begin{center}
\includegraphics[width=120mm] {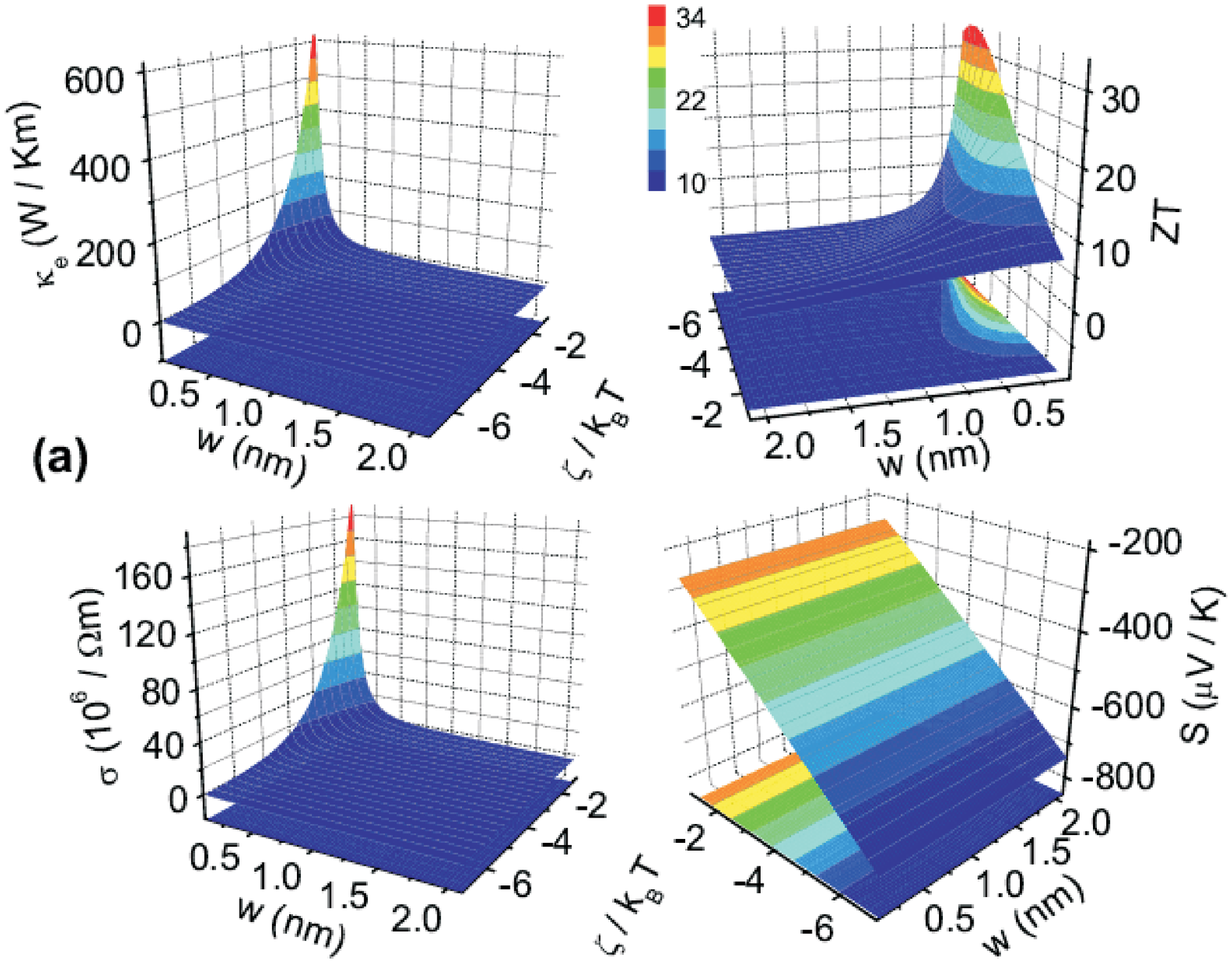}  
\includegraphics[width=120mm] {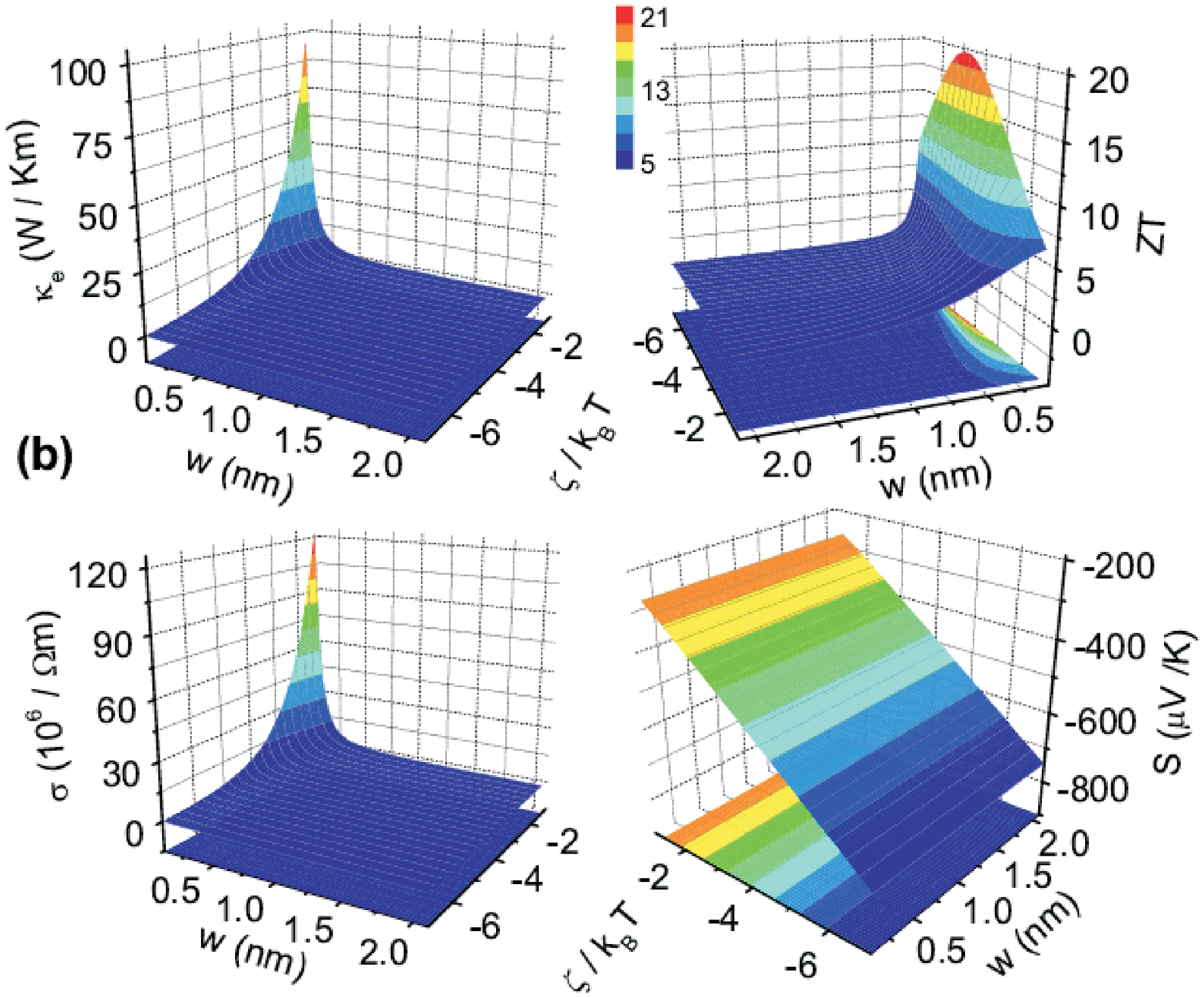} 
\caption{\textbf{(a)} Thermoelectric figure-of-merit $ZT$ for a square Bi$_2$Te$_3$ nanowire at $T=300$ K obtained by the constant-$\tau$ model. The chemical potential $\zeta$ is measured in units of thermal energy $k_{B}T$. The effective mass values are the same as in the Text (${m_x}=0.32$, ${m_y}=0.08$, and ${m_z}=0.02$ along the wire axis). \textbf{(b)}  $T=70$ K. Notice that the predicted enhancement of the transport properties here emerge only in ultra thin wires ($w\lesssim 1$ nm). This has to be compared to the predictions of our model presented in the Text.}
\label{FigSup}
\end{center}
\end{figure} 
%%%%%%%%%%%%%%%%%%%%%%%%%%%%%%%%%%%%%%%%%%%%%%%%%%%%%%%%%%%%%%%%%     
Above, $\varepsilon\equiv{E}/{k_B}T$ is the reduced energy, $m^\ast$ and $\mu_e$ stand for the electron effective mass and mobility  along the wire axis, $\nu$ represents valley degeneracy, and ${\zeta^\ast}\equiv\zeta/{k_B}T$ is the reduced chemical potential of electrons measured from the bottom of the conduction band. The above expressions can  be further simplified to:
%%%%%%%%%%%%%%%%%%%%%%%%%%%    
\begin{eqnarray}
\nonumber
\sigma&=&\frac{\nu{q}}{\pi{w^2}}{\left( \frac{2{k_B}T}{\hbar^2}\right)^{1/2}}{\sqrt{m^\ast}}\mu_{e}F_{-1/2} \\\nonumber 
S&=&-\frac{k_B}{q}\left(\frac{3F_{1/2}}{F_{-1/2}}-{\zeta^\ast} \right)  \\\nonumber 
{\kappa_e}&=&\frac{1}{q}\frac{2\nu}{\pi{w^2}}\left(\frac{2{k_B}T}{\hbar^2}\right)^{1/2}{{k_B^2}T}{\sqrt{m^\ast}}\mu_{e}\left(\frac{5}{2}F_{3/2}-\frac{9F_{1/2}^2}{2F_{-1/2}} \right).
\end{eqnarray}
%%%%%%%%%%%%%%%%%%%%%%%%%%%%%%%  
The dimensionless figure-of-merit, $ZT=\sigma{S^2}T/({\kappa_e}+{\kappa_{ph.}})$, can now be written as: 
%%%%%%%%%%%%%%%%%%%%%%%%%%%%   
\begin{equation}
\nonumber 
ZT=\frac{\frac{1}{2}{\left(\frac{3F_{1/2}}{F_{-1/2}}-{\zeta^\ast}\right)^2}F_{-1/2}}{\frac{1}{B}+\frac{5}{2}F_{3/2}-\frac{9F_{1/2}^2}{2F_{-1/2}}}
\end{equation}
%%%%%%%%%%%%%%%%%%%%%%%%%%%%%%%%% 
with 
%%%%%%%%%%%%%%%%%%%%%%%%% 
\begin{equation}
\nonumber 
B\equiv\frac{2\nu}{\pi{w^2}}\left( \frac{2{k_B}T}{\hbar^2}\right)^{1/2} \frac{{k_B^2}T\sqrt{m^\ast}\mu_{e}}{q\kappa_{ph.}}.
\end{equation}
%%%%%%%%%%%%%%%%%%%%%%%%%%%%%%%%%%%%%%%%         
The most distinct difference of the above expressions  with the those presented in the Text is lack of oscillatory dependence of the transport coefficients on  the wire size and on the quantization of the energy levels, \textit{i.e.}, absence of the quantum size effects. Also, it is noticeable that here the Seebeck coefficient $S$ is wholly independent of size and the temperature, and $\sigma$ and $\kappa_e$ have only a bare monotonic dependence of the form $\propto{1/w^{2}}$ on the wire width. Such dependencies lead ultimately to unrealistically large values for $\sigma$, $\kappa_e$ and, eventually, $ZT$ in ultra thin wires.  Figure \ref{FigSup} presents the obtained results for a Bi$_2$Te$_3$ nanowire with the same parameters as given in the Text.

This work is dedicated to the memory of Amirkhan Qezelli, the uncle, the childhood friend.


\begin{thebibliography}{99}
\bibitem{rhov2} This is not a principal restriction on the present model. If the density of scatterers ${\varrho}$ and their strength $V_0$ (or even only the product $\varrho{V_0^2}$) are known for any single nanowire, one can evaluate all the kinetic coefficients accordingly. The use of a constant mobility here is only to facilitate the comparison to experimental data and to the results of previous models with a constant relaxation time and a fixed mobility.
\bibitem{kappa} This is  a conservative estimation. If one  takes the phonon confinement effects into account, then $\kappa_{ph.}\sim\frac{1}{3}\kappa_{ph.}^{\rm bulk}$ 
(cf.  Ref. \cite{hicks93II}), and one will arrive at a larger value for $ZT$.
\end{thebibliography}
\end{document}